\documentclass[aps,prl,showpacs,twocolumn,superscriptaddress]{revtex4}
\usepackage{amsmath,color}
\usepackage{amssymb}
\usepackage{epsfig}
\usepackage{graphicx,amsmath}
\begin{document}

\title{On the theory of domain switching kinetics in ferroelectric thin films.}
\author{E.V. Kirichenko}
\affiliation{Opole University, Institute of Mathematics and Informatics, Opole, 45-052, Poland}
\author{V.A. Stephanovich}\homepage{http://draco.uni.opole.pl/~stefan/VStephanovichDossier.html} \email{stef@math.uni.opole.pl}
\affiliation{Opole University, Institute of Physics, Opole, 45-052, Poland}
\begin{abstract}
We investigate theoretically the polarization switching kinetics in ferroelectric thin films. In such substances, the
domain walls are pinned by (usually dipole) defects, which are present also in ordered samples as technologically unavoidable impurities. This random interaction with dipole pinning centers results, in particular, in exponentially broad distribution of switching times. Under supposition of low pinning centers concentration, we derive the distribution function of switching times showing that it is not simply Lorentzian (as it was first suggested by Tagantsev et al [\prb, {\bf 66}, 214109 (2002)] ), but is a "square of Lorentzian", which is due to the vector nature of electric field. This improved formalism delivers a better description of available experimental data and elucidates the physical mechanism of domain switching times distribution.
\end{abstract}

\pacs{77.80.Fm, 77.55.-g, 77.84.Lf} \maketitle

Fundamental understanding of polarization switching kinetics in ferroelectric thin films is very important as there are
numerous applications of this phenomenon, the most important being in ferroelectric memories \cite{scott}. In bulk ferroelectric crystals, the successful description of switching kinetics had been achieved using Kolmogorov-Avrami-Ishibashi (KAI) \cite{kol, avr, is1,is2} model focused on the statistics of nucleation of reversed polarization droplets within domains. It has been shown in Ref. \cite{waser} that application of KAI model to ferroelectric thin films does not adequately describe the experiment in the case when wide range of switching times is considered. Tagantsev et al \cite{tag} have shown that this is primarily because the KAI model does not account for broad switching times distribution, which is especially important in the second step of reversed polarization droplets evolution, namely the coalescence of initial droplets into larger (inversely polarized) domains. This is because in KAI model any single droplet of reverse polarization, expanding unrestrictedly, can provide switching of the whole sample. To improve the
description of experimental situation, Tagantsev et al \cite{tag} suggested nucleation-limited switching (NLS) model
which considers a sample as an ensemble of elementary regions which incorporates naturally the broad distribution of their switching times.

While in ceramic or polycrystalline samples this distribution is due to disorder, in crystalline samples there are many unavoidable defects, which also yield the distribution of switching times. It is firmly established (see \cite{tyb} and refs therein), that defects and impurities in ferroelectrics play a role of pinning centers for domain walls so that under external electric field the domain walls motion has a character of creep. This creep motion implies following dependence of a domain wall speed on the local electric field ${\overline E}$ at a pinning site $v \propto 1/\tau \propto$ $\exp[-(U/k_BT)(E_0/{\overline E})^\mu]$, where $\tau $ is a domain wall switching time, $U$ is an energy barrier between different polarization directions and $E_0$ is a critical field for a domain wall creep \cite{tyb,kolton}.

 It has been shown (see, e.g. \cite{tyb} and refs therein), that dynamical exponent $\mu$ (defining the details of a   domain wall pinning potential) almost equals to one so that one can get following expression for field dependent domain wall switching time
\begin{equation}\label{sw1}
    \tau=\tau_\infty \exp\left[\frac{\lambda}{E_{\rm ext}+E}\right],
\end{equation}
where $\lambda=UE_0/k_BT$, $\tau_\infty$ is the switching time at infinite local field ${\overline E}$. Here, following Ref. \cite{lox}, we identify the local field at a pinning site as a sum of external electric field $E_{\rm ext}$ and internal random field $E$ (below we will see that this is actually a modulus of electric field vector) which appears due to disorder.

In non-KAI models of switching, where a distribution of switching times is assumed, the expression for the volume fraction $p(t)$ of a ferroelectric switched by time $t$ assumes the form \cite{tag,lox}
\begin{equation}\label{sw2}
 p(t)\equiv\frac{\Delta P(t)}{2P_s}=\int \limits_{\ln y_{\rm min}}^{\ln y_{\rm max}} \left(1-\exp(-t/\tau)\right)g(\log y)d\log y,
\end{equation}
where $y=\tau/\tau_\infty$, $y_{\rm min}$ and $y_{\rm max}$ are, respectively, the lower and upper limits of the spectrum of dimensionless switching (waiting) times in the system. Their precise form will be determined below. As creep character of domain walls motion implies broad spectrum of switching times, their distribution function
$g(\log (\tau/\tau_\infty))$ is assumed to be broad enough to describe this spectrum.
Its specific form had been chosen as Lorentzian \cite{tag,lox}.

In this Letter, we show that for low pinning centers concentration, when they are spatially uncorrelated (see Refs. \cite{stoneh,anderson} for details), it is possible to calculate exactly the distribution function $f({\bf E})$ of local electric fields ${\bf E}$ at the domain pinning sites. We show that this distribution function reduces to the one of the electric field modulus $E\equiv |{\bf E}|$ and has the form of "square of Lorentzian" rather then simple Lorentzian. This is because pure Lorentzian gives the distribution of a scalar quantity, which in the case of above
dipole defects ensemble corresponds to their resonant frequencies \cite{stoneh,anderson}. At the same time, the
application of formalism \cite{stoneh}  to the vector quantity like electric field gives for its modulus a square of Lorentzian. Having the distribution function $f(E\equiv |{\bf E}|)$ of local electric fields ${\bf E}$ in the domain pinning sites, we can derive the desired distribution function of switching times using the relation \eqref{sw1}. The same expression \eqref{sw1} with respect to the fact that $f(E)$ depends only on modulus $E$, which is a positive quantity, permits to determine uniquely the bounds of switching times spectrum
 \begin{equation}\label{sw2a}
 \tau_{\rm min}=\tau_\infty,\ \tau_{\rm max}=\tau_\infty e^{\lambda/E_{\rm ext}}.
 \end{equation}
The expression for upper bound is consistent with Mertz's law \cite{merz} where the characteristic time of FE polarization switching has the form $\exp(\lambda/E_{\rm ext})$ with $\lambda$ being activation field. In our approach
parameter $\lambda =UE_0/k_BT$ receives theoretical explanation. The expression \eqref{sw2a} also justifies theoretically Eq. (11) of Ref. \cite{tag}, its derivation will be given below.

In the spirit of the statistical method \cite{stoneh}  we write the field distribution function in the form
\begin{equation}\label{sw3}
f({\bf E})=\overline{\delta ({\bf E}-{\bf E}({\bf r_1},...,{\bf r_N}))},
\end{equation}
where bar means the averaging over spatial disorder and ${\bf E}({\bf r_1},...,{\bf r_N})$ is the electric field created by
$N$ randomly distributed defects. At low pinning centers concentration, where their spatial correlations (which means, for example, the absence of close pairs of defects) are small, the contributions of various defects to ${\bf E}$ are simply additive \cite{stoneh}
\begin{equation}\label{sw4}
{\bf E}({\bf r_1},...,{\bf r_N})=\sum_{i=1}^{N}{\bf E}({\bf r_i}).
\end{equation}
The dipole nature of defects implies that
\begin{equation}\label{sw5}
{\bf E}({\bf d},{\bf r})=\frac{d}{r^3}\left[3({\bf l}\cdot {\bf n}){\bf n}-{\bf l}\right],
\end{equation}
where ${\bf l}\equiv {\bf d}/d$ and ${\bf n}={\bf r}/r$ are unit vectors directed along ${\bf d}$ (vector of dipole moment) and ${\bf r}$ respectively. Using integral representation of 3D delta-function \eqref{sw3} and factorizing it by Eq. \eqref{sw4} with respect to \eqref{sw5}, we obtain following expression for distribution function
\begin{equation}\label{sw6}
f({\bf E})\equiv f(E)=\frac{1}{(2\pi)^3}\int e^{\imath{\bf E}{\mathbf x}-\delta x} d^3x,\ \delta=\frac{4\pi^2}{9\sqrt{3}}nd.
\end{equation}
Here $n=N/V$ is pinning centers (dipoles) concentration, $d$ is its dipole moment. The distribution function \eqref{sw6} depends on vector ${\bf E}$ modulus due to spherical symmetry of scalar products. The evaluation of integral \eqref{sw6} gives following explicit expression for field distribution function
\begin{equation}\label{sw7}
f(E)=\frac{\delta}{\pi^2(E^2+\delta^2)^2}, \ 4\pi\int_0^\infty E^2f(E)dE=1.
\end{equation}
The expression \eqref{sw7} for $f(E)$ along with Eq. \eqref{sw1} permits to obtain the explicit form of the switching times distribution function $g(\ln y)$, $y=\tau/\tau_\infty$. For that we demand that function $g(\ln y)$ should be normalized to unity, namely
\begin{eqnarray}\label{sw7a}
&&4\pi \int_{E(\psi)=0}^\infty E^2(\psi)f(E(\psi))\frac{dE(\psi)}{d\psi}d\psi=\nonumber \\
&&=\int_{\psi_{\rm min}}^{\psi_{\rm max}}g(\psi)d\psi=1, \ \psi=\ln y \equiv \frac{\lambda}{E_{\rm ext}+E}.
\end{eqnarray}
With respect to the fact that $dE/d\psi<0$, Eq. \eqref{sw7a} implies that $\psi_{\rm min}=\psi(E\to \infty)=0$ and
$\psi_{\rm max}=\psi(E=0)=\lambda/E_{\rm ext}$. This gives the above expression \eqref{sw2a}.

We emphasize here that for ordinary Lorentzian (we can obtain it supposing that electric field is a scalar quantity) the natural boundaries of switching times spectrum $\tau_{\rm min}$ and $\tau_{\rm max}$ do not appear - both $E$ (now it is a scalar) and $\ln \tau$ can be varied from $-\infty$ to $\infty$. We also note that the case $\ln \tau =-\infty$
is unphysical as it means zero switching time, i.e. system switches its polarization immediately. This is impossible as
dipoles have finite mass so that some time should pass for polarization to reverse its direction. This means that consistent account for vector nature of electric field gives physically correct switching times spectrum in a ferroelectric.

The explicit form of $g(\psi)$ can easily be extracted from Eq. \eqref{sw7a} to read

\begin{eqnarray}\label{sw10}
&&g(\psi)=\frac{4\lambda \delta}{\pi}\frac{(\lambda-\psi E_{\rm ext})^2}{\left[(\lambda-\psi E_{\rm ext})^2+\delta^2\psi^2\right]^2}, \nonumber \\
&&\int_0^{\lambda/E_{\rm ext}}g(\psi)d\psi=1.
\end{eqnarray}
The expression \eqref{sw10} gives the desired distribution of switching times as a function of pinning centers (dipoles) concentration, temperature and external electric field. Note that distribution function \eqref{sw10} depends also on the microscopic parameters of a material like pinning center dipole moment $d$ as well as on parameters $E_0$ and $U$ of domain walls subsystem.

\begin{figure} [! ht]
\begin{center}
%%\vspace*{-0.8cm}
\includegraphics [width=0.49\textwidth]{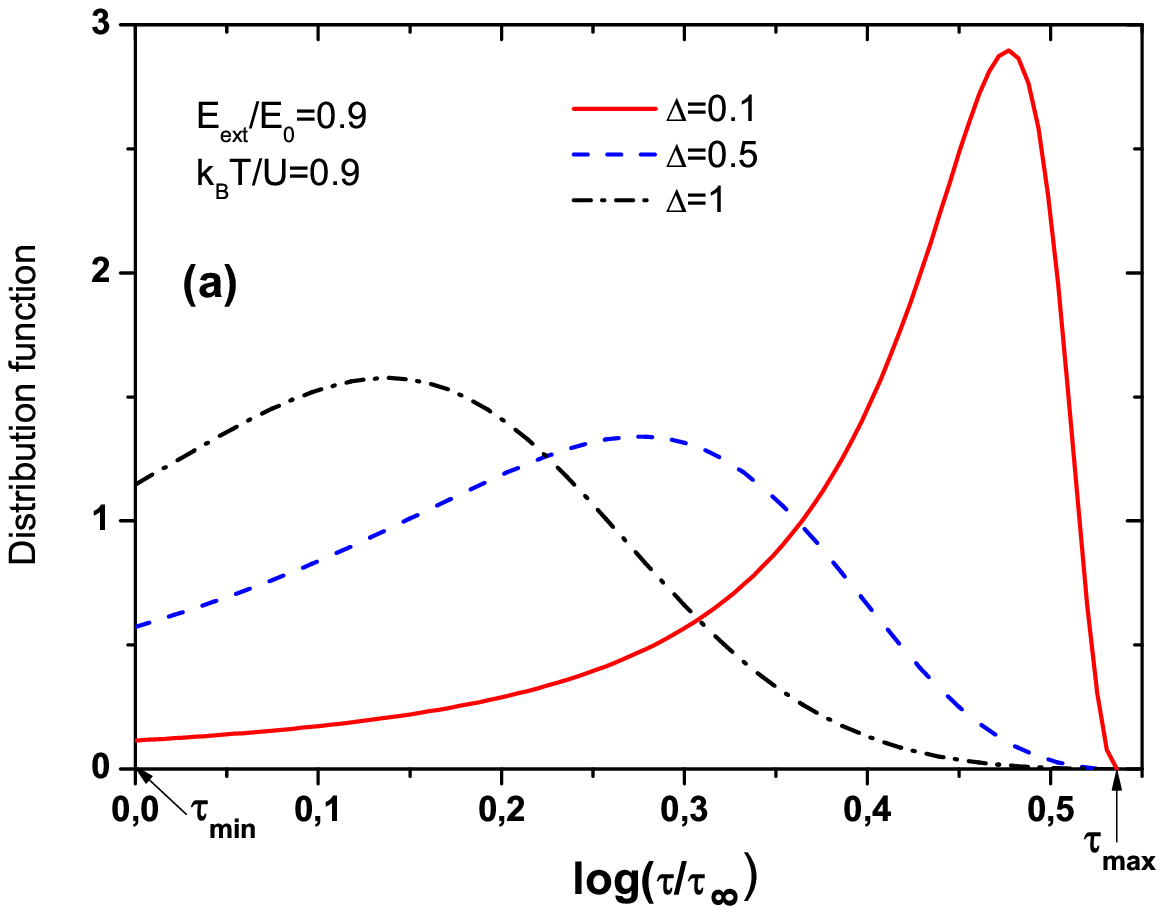}
\includegraphics [width=0.49\textwidth]{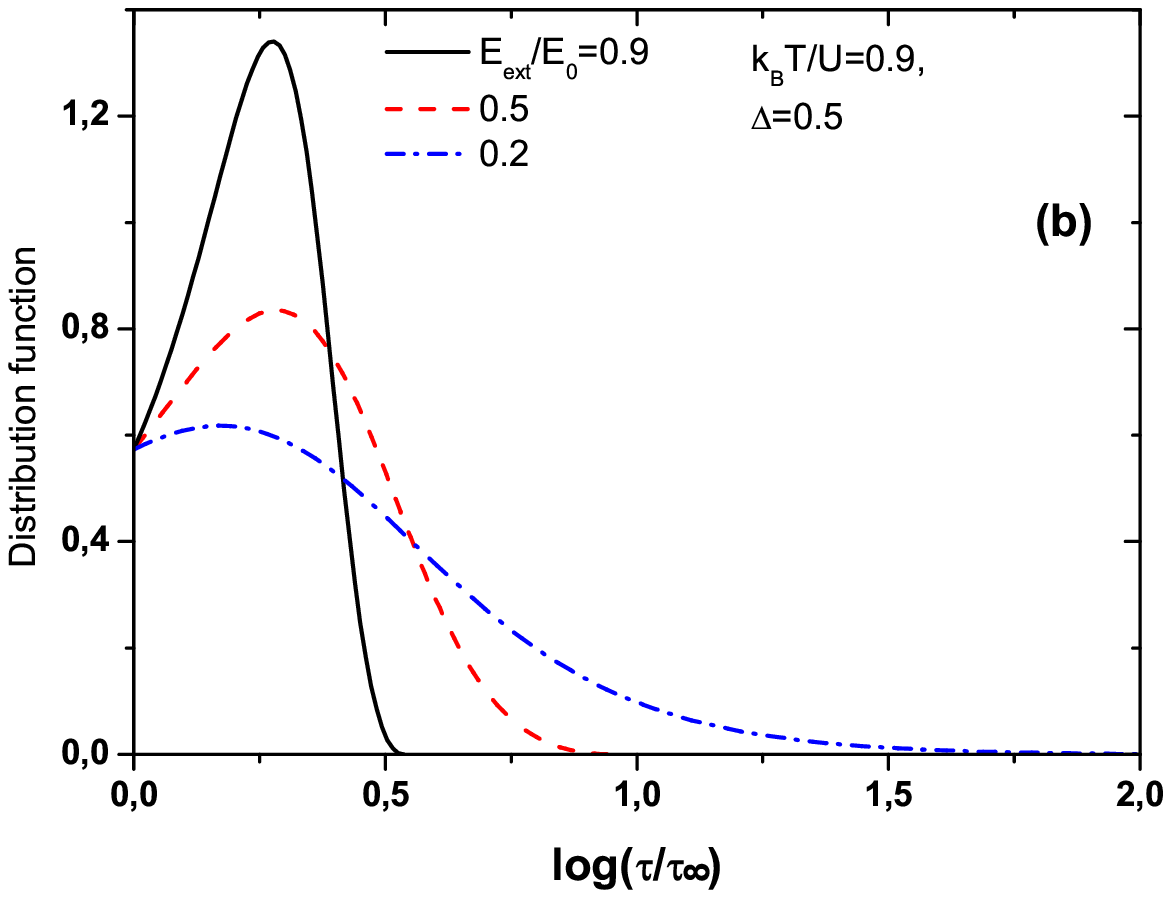}
\includegraphics [width=0.49\textwidth]{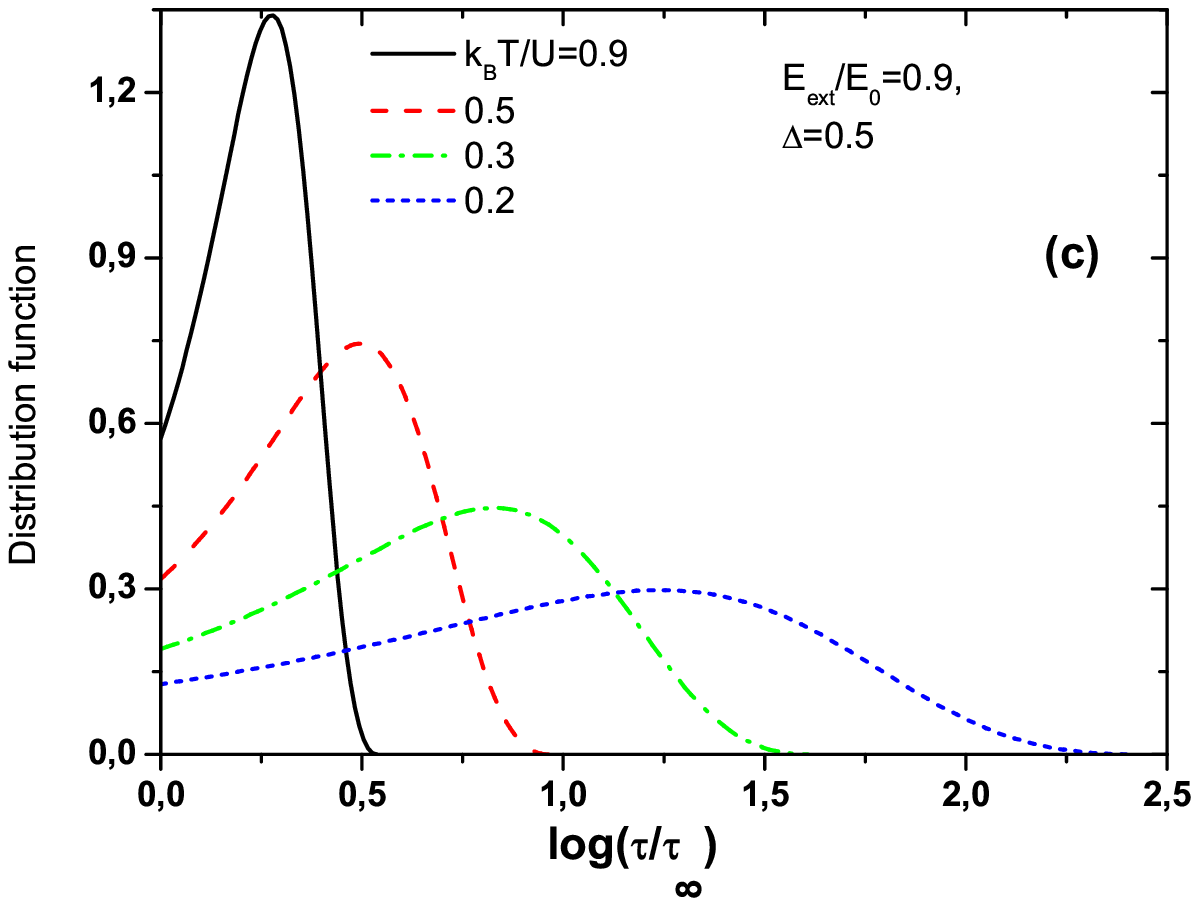}
\end{center}
\vspace*{-0.8cm}
\caption{Distribution function of switching times as a function of: dimensionless width (product of dipoles concentration and its dipole moment) (a),
electric field (b) and temperature (c). Note the common logarithms on the abscissa axis.}\label{ufi1}
\end{figure}

The plots of $g(\psi)$ (in dimensionless variables $\Delta =\delta/E_0$ $=\frac{4\pi^2}{9\sqrt{3}}\frac{nd}{E_0}$, $\eta=k_BT/U$, $\xi=E_{\rm ext}/E_0$) are reported on Fig.\ref{ufi1} at different fields, temperatures and $\Delta$'s. It is seen from Fig.\ref{ufi1}a (where $\tau _{\rm max}$ and $\tau _{\rm min}$ are also shown, which are constants in this case) that at small widths $\Delta$ the distribution function is peaked near $\tau _{\rm max}$, becoming delta function as $\Delta \to 0$. This means that at small pinning centers concentration the systems switched nonrandomly with the same time $\tau _{\rm max}$, which obeys Mertz's law \eqref{sw2a} \cite{merz}. In the opposite limiting case the degree of disorder is maximal so that the distribution function is spread evenly between $\tau _{\rm min}$ and $\tau _{\rm max}$. Fig.\ref{ufi1}b shows the field dependence of the distribution function at fixed temperature and $\delta$. According to the physics of our system, at high fields the distribution function is localized near $\tau_\infty=\tau _{\rm min}$ (note that at high fields $\tau _{\rm max} \to \tau _{\rm min}$, see Eq. \eqref{sw2}), while at lower fields the distribution function smears between $\tau_\infty=\tau _{\rm min}$ and $\tau _{\rm max}$ which grows infinitely at field lowering. Thus, the electric field also effectively inhibits disorder in the system - at high fields the switching occurs at the same time $\tau_\infty$. Note that this conclusion coincides with the results of Ref. \cite{tag}. Qualitatively similar behavior is reported on Fig.\ref{ufi1}c, where the temperature dependence of $g(\psi)$ \eqref{sw10} (at fixed field and $\delta$) is depicted. At hight temperatures the distribution function is localized near $\tau _{\rm min}$, while at low temperatures it becomes delocalized. This is because in the switching process there is a competition between pinning and thermal energies of a dipole. At sufficiently high temperatures the thermal energy $k_BT$
exceeds the majority of energies of pinning centers so that the width of $g(\psi)$ in this case can be regarded as a measure of number of "turned on" pinning centers. Take it other way, at low temperatures, the random pinning in the system is strong and hence the distribution function is wide, while
at high temperatures almost all pinning centers are "turned off" (their energies are smaller then the energies of dipoles thermal reorientations) and distribution function is narrow. Thus, at sufficiently high fields and/or temperatures the system becomes effectively ordered and switching occurs coherently in time at $\tau = \tau_\infty=\tau _{\rm min}$.

The expression \eqref{sw10} can be readily substituted into \eqref{sw2} to calculate switching polarization fraction $p(t)$.  Latter calculation, however, can be done more conveniently in terms of field variable $E$. Namely, $<e^{-t/\tau}>=4\pi\int_0^\infty E^2 f(E) e^{-t/\tau(E)}dE$, where $\tau(E)$ is given by Eq. \eqref{sw1}. Proceeding along these lines we obtain following expression for $p(t)$
\begin{equation}\label{sw11}
p(t)=1-\frac{4}{\pi}\int_0^\infty \frac{z^2dz}{(z^2+1)^2}\exp\left[-t_1e^{-\frac{1}{\eta(z\Delta+\xi)}}\right],
\end{equation}
where $t_1=t/\tau_\infty$. Note that similar to Refs.\cite{tag, cors} we can pass to logarithmic variables and change exponential function in the integrand \eqref{sw11} to unit step function centered at $z_0=(1/(\eta \ln t_1)-\xi)/\Delta$.
This gives an analytical approximation to the integral \eqref{sw11} \cite{an}. However, this analytical approximation is not always accurate.

\begin{figure} [! ht]
\begin{center}
%%\vspace*{-0.8cm}
\includegraphics [width=0.49\textwidth]{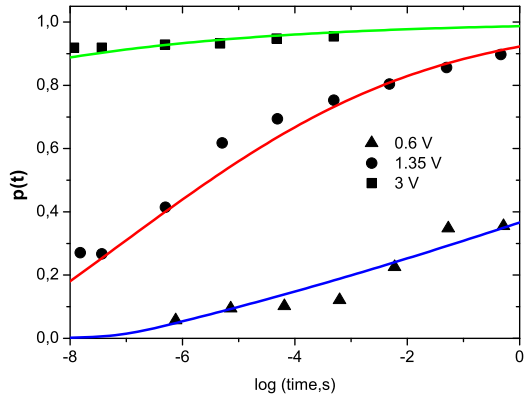}
\end{center}
%\vspace*{-0.8cm}
\caption{Comparison between our theoretical switching curve \eqref{sw11} with experimental points from Ref.\cite{tag}.}\label{ufi2}
\end{figure}

The results of numerical calculation of $p(t)$ \eqref{sw11} are reported on Fig.\ref{ufi2}. On the same figure, as an illustrative example, we plot the experimental points from Ref. \cite{tag}, where the switching kinetics in Pt/Pb(Zr,Ti)O$_3$/Pt thin films had been measured. It is seen, that distribution function \eqref{sw10}, similarly to simple Lorentzian, provides a good coincidence between theory and experiment. As function \eqref{sw10} decays at infinities like $1/\ln^2(\tau/\tau_\infty)$ (this decay character is similar to Lorentzian), the switching curve \eqref{sw11} covers many decades in $\log(t)$ (here we consider the common logarithm of time).

Our analysis shows, that expression \eqref{sw11} for $p(t)$ can (at different values of parameters $\Delta$, $\eta$ and $\xi$) describe also the experiments from Refs.\cite{lox,gruv}. This means that the above approach is at least not worse then previous descriptions based on the phenomenological introduction of Lorentzian distribution. For our fits, we take $\tau_\infty=10^{-12}$ s, $\eta=0.2$ (the measurements in Ref.\cite{tag} had been performed, most probably, at room temperature) and $\xi=0.07$, $\Delta=0.07$ for voltage $v=0.6$ V, $\xi=0.1$, $\Delta=0.2$ for $v=1.35$ V and $\xi=0.3$, $\Delta=0.85$ for $v=3$ V. The fact that we need larger $\Delta$ for fitting at higher fields is
consistent with our theory as at increase of $\Delta$ the maximum of distribution function shifts towards $\tau_\infty$, see Fig.\ref{ufi1}a. The higher fields, by themselves, make the system to switch coherently at smallest possible time $\tau_\infty$, which is seen on the Fig. \ref{ufi2}, where the $p(t)$ curve for $v=3$ V has been switched already at the times less then $10^{-8}$ sec. The electric field may increase $\Delta$ via the dipole moments of pinning centers or their interaction with domain walls. Theoretical explanation of this fact is out of frames of present paper. Actually, to have reliable base for creation of microscopic theory of domain nucleation and switching in thin ferroelectric films, the detailed experimental investigations of microscopic characteristics of domain walls motion are highly desirable.

To conclude, we have shown that our model based on "squared Lorentzian" distribution function gives a good description of the switching kinetics in ferroelectrics thin films. Generally speaking, our treatment has to do with disordered ferroelectric films, where long-range random interaction is responsible for distribution of switching times. Even in ordered (epitaxial) films there are randomly situated technologically unavoidable defects (like stacking faults in PZT films, \cite{tag}), which could be the source of above distribution function of random fields and hence of switching times. We note here, that in perfectly ordered films the theory of ferroelectric switching may be constructed on the base of continuous Chensky and Tarasenko approach \cite{ct} (see Eq. (38) of Ref. \cite{ct}),augmented by relaxation term. Namely, the equation of motion for the polarization vector ${\bf P}$ should have the form $\mu \ddot{{\bf P}}+\eta \dot{{\bf P}}=-\delta F/\delta {\bf P}$, where $F$ is corresponding (nonlinear) Ginzburg-Landau free energy including polarization gradients and, if necessary, the contribution from electrodes. This equation of motion should be augmented by the Maxwell equations for polarization outside the sample and proper boundary conditions accounting, in particular, for depolarization field. This problem is surely doable numerically, but for thin films it is doable analytically as domain structure there is always "sinusoidal" \cite{lev}, i.e. it is described by the linear equation od state $f({\bf P})=-\delta F/\delta {\bf P}=0$. However, the proper account for disordered ensemble of pinning centers seems to be problematic in such continuous theory. That is why the approaches based on the "improvement" of KAI model by the introduction of nucleus switching time distribution seem to be reasonable for not perfectly ordered ferroelectric films.

We are grateful to B.A. Ivanov for enriching discussions.

\end{document}